# BioNet-XR: Biological Network Visualization Framework for Virtual Reality and Mixed Reality Environments

Busra Senderin[1], Nurcan Tuncbag[2,3], and Elif Surer[1,*]

[1] Department of Modeling and Simulation, Graduate School of Informatics, Middle East Technical University, Ankara, 06800, Türkiye

[2] Department of Chemical and Biological Engineering, Koç University, Istanbul, 34450, Türkiye

[3] School of Medicine, Koç University, Istanbul, 34450, Türkiye

* To whom correspondence should be addressed.

**Abstract**

Protein-protein interaction networks (PPIN) enable the study of cellular processes in organisms. Visualizing PPINs in extended reality (XR), including virtual reality (VR) and mixed reality (MR), is crucial for exploring subnetworks, evaluating protein positions, and collaboratively analyzing and discussing on networks with the help of recent technological advancements. Here, we present BioNet-XR, a 3D visualization framework, to visualize PPINs in VR and MR environments. BioNet-XR was developed with the Unity3D game engine. Our framework provides state-of-the-art methods and visualization features including teleportation between nodes, general and first-person view to explore the network, subnetwork construction via PageRank, Steiner tree, and all-pair shortest path algorithms for a given set of initial nodes. We used usability tests to gather feedback from both specialists (bioinformaticians) and generalists (multidisciplinary groups), addressing the need for usability evaluations of visualization tools. In the MR version of BioNet-XR, users can seamlessly transition to real-world environments and interact with protein interaction networks. BioNet-XR is highly modular and adaptable for visualization of other biological networks, such as metabolic and regulatory networks, and extension with additional network methods.

**Availability and implementation:** The code and data files are available at: https://github.com/BioNetXR/Release

The supplementary video files are available at Dropbox and the shortened link is: http://bit.ly/42w8zHz

**Contact:** elifs@metu.edu.tr

## 1. Introduction

Proteins interact with each other to perform specific functions. The complete set of protein interactions constitutes the interactome, which is spatially and temporally dynamic (reviewed in (Keskin, et al., 2016)). Network visualization of the interactome is fundamental for better perception of connections between proteins and exploring hidden patterns among them (Shannon, et al., 2003). There are several network visualization tools (Bastian M., 2009; Hu, et al., 2008; Shannon, et al., 2003). Cytoscape is one of the widely used open-source network visualization and analysis programs (Shannon, et al., 2003). Besides 2D visualization, Cytoscape supports 3D visualization via additional plugins like Cy3D. VisANT (Hu, et al., 2008), BioLayout (Wright, et al., 2014), OmicsNet (Zhou and Xia, 2018) and Arena3D (Kokoli, et al., 2023) provide 3D network representation. VMD (Humphrey, et al., 1996) and PyMol (Schrödinger, 2020) are among the most frequently used ones, especially in the 3D visualization of biomolecules. However, these tools do not inherently support 3D stereoscopic models while allowing 3D visualization with high-quality screen displays. Thus, virtual reality (VR) and mixed reality (MR) technologies are increasingly crucial in domains like education and medicine, and their application in life sciences is becoming indispensable. Adaptation of biomolecular structure visualization in VR and MR environments was faster where several tools including Molecular Rift (Norrby, et al., 2015), and ProteinVR (Cassidy, et al., 2020) collaborate in computer and virtual reality to visualize protein structures or explore the protein structure in a real-world setting, i.e., in a room. Another tool, BioVR, delivers a VR experience for the visual examination of DNA, RNA, and protein sequences alongside structural visualization. ConfocalVR (Stefani, et al., 2018) and Juicebox VR (Durand, et al., 2016) demonstrate the use of VR in various projects.

iCAVE (Liluashvili, et al., 2017) and VRNetzer (Pirch, et al., 2021) are among the network visualization tools that provide a 3D interactive environment for visualization and analysis. They allow users to query built-in databases, use a network layout algorithm, and introduce known 2D network extensions to 3D. In iCAVE, software portability was once cited as a primary limitation of VR. Nevertheless, due to the continuous advancements in VR technology, this concern has become less relevant. Additionally, the cost of having low-end and high-end VR headsets has significantly decreased. Another advantage of these tools is that the users do not require a specialized environment. Network visualization can be performed anywhere, whether within a room or in an isolated virtual environment. In very recent work, VRNetzer enables interactive exploration of complex networks and identifies disease-associated genes within molecular networks in a VR platform (Pirch, et al., 2021).

Despite the availability of tools such as VRNetzer, which provides a VR-based approach, and non-immersive tools like iCAVE, the comprehensive integration of biomolecular interactions within virtual reality (VR) and mixed reality (MR) environments for the visualization and analysis of biological networks remains an area yet to be fully addressed in the available literature. In this study, we present BioNet-XR, a PC-, VR-, and MR-based biological network visualization environment. BioNet-XR provides several analysis and visualization features including teleportation between nodes, general and first-person view to explore the network, subnetwork construction via PageRank, Steiner tree, and all-pair shortest path algorithms. Furthermore, we gathered user feedback from both specialists (bioinformaticians) and generalists (multidisciplinary groups), addressing the need for usability evaluations of VR-based visualization tools. Within the desktop and VR versions, users have access to various analytical tools, including simple or shortest path finding and subnetwork discovery while in the MR version, users can seamlessly transition to real-world environments, interacting with protein interaction networks by exporting the networks initially created in the desktop version.

## 2. Materials and Methods

### 2.1. Overview of BioNet-XR

BioNet-XR is developed for biomolecular network visualization where XR represents VR and MR versions together. BioNet-XR works in four steps: the data processing step, the graph data structure construction step, the network visualization step, and the user actions step (**Figure 1**). The first step is reading and parsing the network file that was uploaded by the user so that the nodes (protein names) and edges (interactions) are obtained. A graph structure is created based on this information. In the network visualization, nodes and edges are visualized as spheres and pipes, respectively, and the selected layout and visual settings are applied. In this way, the network creation process is completed. Responding to the user actions involves executing modifications and functionalities after the network is created.

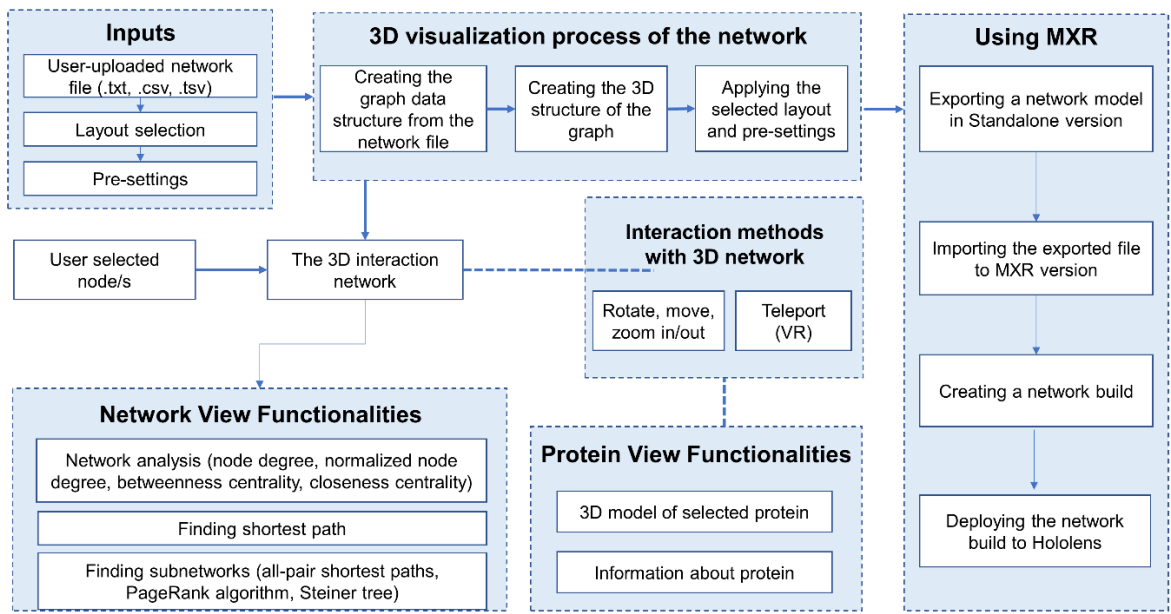

**Figure 1.** General overview of BioNet-XR for 3D network visualization.

BioNet-XR was developed using the Unity3D game engine with the C# programming language. Unity3D supports development in 2D and 3D via cross-platform support. It also offers the possibility to use plugins and toolkits for other disciplines such as artificial intelligence, VR, and AR. The desktop version of BioNet-XR works on both Windows and Mac, the VR version works on HTC Vive, and the MR version on Microsoft Hololens.

In the VR version, HTC Vive should be connected to the computer when developing an application, and it does not provide an emulator. Hence, the Vive Input Utility was used for development and emulator support.

Some system features require an MR application compatible with Microsoft Hololens, a lightweight and affordable headset. The Mixed Reality Toolkit is used to develop applications for the Hololens which is available for Unity3D.

After a network is loaded in the desktop version, the 3D network model and associated data (positions, colors, and sizes of the spheres and lines representing the nodes and edges) is saved in a JSON format (press E key). The next step is exporting from the desktop version and loading it to the MR module which creates the visualization of the same 3D network model with the same layout in MR.

The desktop version is executable without installation. Two viewer modes are provided for the user to examine the network: 'Default Viewer' and 'First-person Viewer.' In the Default Viewer mode, the user can examine the network from the classical viewpoint. Furthermore, the user can move back and forth within the network with the W, S, or arrow keys.

The VR version has been developed to make the BioNet-XR more immersive. It has 3D viewing and almost all the features offered by the desktop version. The head-mounted display (HMD) enables the user to rotate their head, granting a full 360-degree view. The user can navigate the network by teleporting to the nodes with hand controls. Teleporting enables the user to examine the network from many viewing angles that the PC version cannot offer. In the VR version, three different virtual environment theme options are available. The user can choose a theme initially and use it in the selected virtual environment.

The most crucial feature of mixed reality (MR) is supporting the real-world environment instead of a completely virtual environment where the users view the created network in a real-world setting. MR is a newly emerging technology in visualization tools, and there are very few studies available. The MR version which works with the Microsoft Hololens has also been developed to display protein networks by considering its capability.

### 2.2. Subnetwork Identification

All-pairs shortest path means the shortest path between all nodes in the network and is solved with the Floyd-Warshall Algorithm. The weight of the edge from vertex $i$ to vertex $j$ is denoted as

*w(i,j)*. The algorithm computes a matrix *D*, where *D[i][j]* represents the shortest distance from vertex *i* to vertex *j*. For *k* = 1 to the number of vertices in the graph:

$$D[i][j] = \min(D[i][j], D[i][k]+D[k][j])$$

It updates *D[i][j]* by considering all vertices *k* as potential intermediate vertices on the shortest path from vertex *i* to vertex *j*. After *n* iterations, the matrix *D* contains the shortest distances between all pairs of vertices. When finding a subnetwork, only the nodes given as input are taken into account, and the shortest paths between them create the subnetwork. Non-seed nodes on the path between seed nodes are part of the network.

A Steiner tree is a problem that aims to find a tree containing a specific node subset in a graph and is solved by Kou's Algorithm. Assume a graph *G=(V,E)* is the connected, undirected graph, where *V* is the set of vertices and *E* is the set of edges. The weight of an edge *e* is denoted by *w(e)*. At the beginning, the set of edges of the minimum spanning tree *T* is empty.

Until *T* forms a spanning tree (i.e., until *|T| = |V| - 1*), it finds the edge *e* with the smallest weight that does not create a cycle when added to *T* and *e* to *T*. When the iterations end, it returns *T* as the minimum spanning tree of *G*. The algorithm first finds a minimum spanning tree in the given graph. Then, unlike the All-Pairs Shortest Path, it removes other nodes until all remaining leaf nodes are seed nodes.

Random Walk enables finding subnetworks by scoring signals of nodes. A certain amount of signal (fluid, heat) is spread from the seed nodes in the network. An equal signal is transferred to the neighbors in each iteration.

Assume a graph *G=(V,E)* is the connected, undirected graph, where *V* is the set of vertices and *E* is the set of edges. It starts from a chosen node $v_0 \in V$. A function *Step(v)* is defined to select a neighboring node of *v* at each step. This function can be probabilistic, deterministic, or uniform, depending on the type of random walk. The step function is repeated for a fixed number of steps $t$ or until a stopping criterion is met. The sequence of nodes visited during the walk as $v_0, v_1, v_2, …, v_t$ represents the path. In BioNet-XR, users can set the iteration number. The amount of signal on the node at the end of the iterations is its score and the nodes are colored based on their scores. A gradient color palette is used corresponding to score ranges. Users can edit this color palette.

### 2.3. Evaluation of Usability

The International Organization for Standardization's usability definition highlights the importance of enabling users to perform targeted work effectively and satisfactorily (Abran, 2003). Usability is a criterion that is frequently considered and evaluated in software development. Thus, usable software also has a smoother learning curve.

Most of the studies on visualization tools lack the evaluation of usability. BioNet-XR however, fills this gap by being a user-friendly software for use in clinical research and educational settings, capitalizing on the suitability of immersive technologies for collaborative purposes. Therefore, it should be applicable to both the target and general audiences. The tests were conducted with both audience samples. Only the usability tests of the PC version were performed.

### 2.4. Participants

Subjects were invited via email to the online tests and 30 people participated. Among participants aged 24-40, 14 (46.67%) were female and 16 (53.3%) were male. Almost half (46.7%) were MSc candidates, while the remaining (53.3%) had an MSc graduate and further education. 66.6% of the multidisciplinary group was from Multimedia Informatics. While the Bioinformatics group was familiar with the visualization tools and target audience, the multidisciplinary group represented a general audience.

In the open-ended questions, the participants were asked about their previous usage of visualization tools, which ones they used, and how frequently. Out of the total participants, 43.3% (N = 13) stated using visualization tools, while 56.7% (N = 17) did not. While most in-field participants used such a tool, most out-of-field participants did not, as expected. The commonly reported software included Cytoscape, Pymol, Gephi, and NeoVis. Among these, Cytoscape was the most frequently used software based on participant responses.

Also, the participants were asked about their previous experience with VR/MR applications. Due to the pandemic, participants could not test the BioNet-XR's VR and MR versions. For this reason, it was crucial for them to at least watch the usage videos in test sessions to have an idea. In open-ended questions, participants were asked their thoughts on the potential usefulness of the VR/MR versions based on the videos they watched.

### 2.5. Procedure

The procedure was administered online via the Zoom application. Each online test session was scheduled individually for participants on their available date and time. Before the tests, download links for the framework and user guide were provided to the subjects. It was not asked to use the framework or read the user guide. A brief description of the test procedure was described at the beginning of the test. During the test session, the participants could use the user guide and ask for help where they got stuck and shared screens for observation. The test sessions were recorded for later analysis when the participant was allowed.

Subjects were asked to follow four steps in a specified exercise with given data files. An exercise was necessary to prevent non-domain users from feeling lost when using the framework. The completion time of the exercise varied depending on the participant. The most extended test duration was 17 minutes 36 seconds, and the shortest time was 1 minute 60 seconds. The average completion time was 5 minutes and 13 seconds.

After the exercise was completed, they watched the videos of VR and MR versions to get an idea. Finally, they were asked to fill out online questionnaires. The tests took approximately one hour, including completing the exercise, watching the videos, and completing the questionnaires.

### 2.6. Exercise Design

In the task-based method, the user is asked to complete specific steps. It is widely used in testing visualization tools, which was also applied in this evaluation. It shows that the participants did not feel lost, and the usability tests were efficient. Hence, an exercise was prepared in such a way as to ensure that each participant tests at least the framework's following primary functions: creating, controlling, manipulating networks, and finding short paths and subnets.

It is also essential that the prepared task be as close to an actual use case. Hence, the case study in the exercise was designed based on Chen et al. (Chen, 2019) constructing and analyzing PIN for heroin addiction. The study provides information about a heroin addiction network. Such a case study helps non-experts in comprehending the software's purpose and its intended application. Additionally, some helpful information and definitions were given in the exercise. Further, without the need to answer a question or make inferences, users can focus only on the framework without feeling pressured or stressed.

A data file suitable for BioNet-XR was created using the provided information in the study on which the exercise was based and given to the participants at the beginning of the test. Thus, the participants reproduced the network in 3D in BioNet-XR. Additionally, the study shows common proteins that are also effective in alcohol, cocaine, and amphetamine addictions. Appropriate data files were prepared with these proteins for seed node selection while finding subnetworks in BioNet-XR. It directs participants to find subnets of proteins effective in heroin and other addictions using the provided file.

### 2.7. Data Collection Instruments

The instruments are the System Usability Scale (SUS), Technology Acceptance Model (TAM), and open-ended questions. John Brooke introduced the System Usability Scale (SUS) in 1996 to measure the usability of systems as context-independent, which makes it easy to use and flexible (Brooke, 1996). It is a 5-point Likert scale (1=Strongly Disagree, 5=Strongly Agree) with ten items. Bangor et al. presented adjective ratings to make SUS scores more understandable (Bangor, 2009). Seven adjectives correspond to the score ranges in the adjective scale, and from the lowest usability rating to the highest, these adjectives are: worst imaginable, awful, poor, ok, good, excellent, and best imaginable.

The Technology Acceptance Model (TAM) (Davis, 1986) is a framework for assessing information systems. It enables an understanding of the target audience's technology acceptance level of a system. It consists of various assessments that measure multiple aspects of how an information system is used. The questionnaire consists of 14 questions and is scaled between 0 and 10 (0 = Strongly Disagree, and 10 = Strongly Agree).

Open-ended questions were prepared by considering the questions in similar studies (Davis, 1986; Javahery, 2007; Marcus, 2005). It consists of questions about participants' previous experiences with VR and MR technologies and their impressions of BioNet-XR's versions.

### 3. Results and Discussion

When the user starts BioNet-XR, two options are given: PC-version and VR-version. VR-version has three themes (**Video S1**). For illustration purposes, we show the Glioblastoma (GBM) network from the reference (Tuncbag, et al., 2016) which has 83 nodes and 106 edges (**Figure 2a**, **Video S2**). During visualization, users have the flexibility to explore the network from a general perspective or choose a specific node for a first-person perspective exploration (**Figure 2b**, **Video S3**). First-person viewer mode is an integrated version of the first-person camera used in video games. The difference between the default mode and this mode is the camera's mobility. In default mode, the camera is fixed, and the network has mobility. The network rotates and moves closer and further away from the camera. In the first-person viewer mode, the network no longer needs to move because the camera is like a placeholder for the user and has mobility. The user rotates the head and navigates through the network as the camera. Users can find the shortcut between the two nodes and find the subnetwork with nodes selected. They can perform various modifications on the network, such as selecting the node, moving the node, and viewing neighbor nodes.

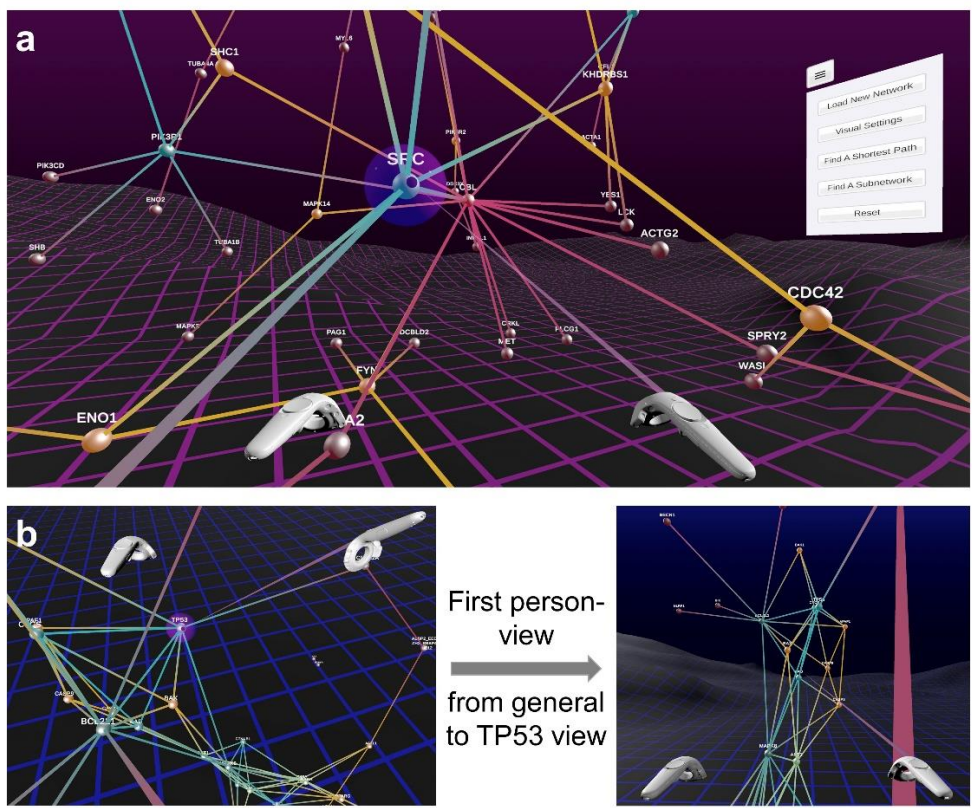

**Figure 2.** Representative network visualization in VR (a) General view with a selected node, (b) switching from a general view to a first-person view for TP53.

BioNet-XR can compute network parameters, determine the shortest path between two nodes, and find subnetworks from a list of genes by providing a range of algorithm options. Files in the local source can load into the application via a file browser. Users can upload files with .txt, .csv, or .tsv extensions where each line represents the connection between protein pairs. The comma character separates protein names. A network file given as input is simply a list of protein links. Layout and visual settings are configured during the network file upload phase. Visual settings consist of settings such as font size, node size, and color. The appearance of the representations is tunable with these parameters.

A graph data structure is created by parsing the selected network file. Each protein name corresponds to a node, and the link between two proteins corresponds to an edge. When the operation is completed, nodes (proteins) are visualized as spherical objects, and edges (bonds between proteins) as lines between two related nodes. It provides visual settings such as assigning the size and color of the node according to the node degree and adjusting the font size of the node names.

When clicking on a node, incident edges and adjacent nodes are highlighted in red, and the edges are in pink. Its network parameters are shown such as node degree, closeness centrality, and betweenness centrality.

In graph visualization, it is needed to place nodes and edges properly to get a clear and understandable view. The layout algorithms are used to determine how a network will lay out in the space. Offering more than one layout is a common feature of visualization tools. BioNet-XR includes three layouts to show the placement of nodes and edges neatly and aesthetically (**Video S4**, **Video S5**, **Video S6**, **and** **Video S7**).

In the force-directed layout, the graph is placed by modeling the nodes as if there is a physical push-pull force between them. Fruchterman and Reingold [15] developed an algorithm for 2D graphs adapted to 3D. Nodes are assigned random positions initially. The push-pull force on the nodes is calculated in each iteration, and how much the node will move is computed considering the temperature value, which affects the maximum amount the node can move and this amount decreases over time. The force-directed layout is commonly included in most visualization tools of biological networks [16].

The single circular layout is based on the barycenter heuristic [17]. Implementing a barycenter heuristic allows nodes to be placed as close as possible to the nodes they are connected to, rather than randomly around a circle. The combination of Louvain Community Detection and Single Circular Layout is presented as a third layout. First, the clusters in the network are determined with Louvain Community Detection. Then, the single circular layout is applied to the clusters, and the final layout of the network is obtained.

The shortest path between a start node and an end node is found using the breadth-first search algorithm. When a route is found, nodes as red and edges as pink are visualized (**Figure 3**, **Video S8** and **Video S9**).

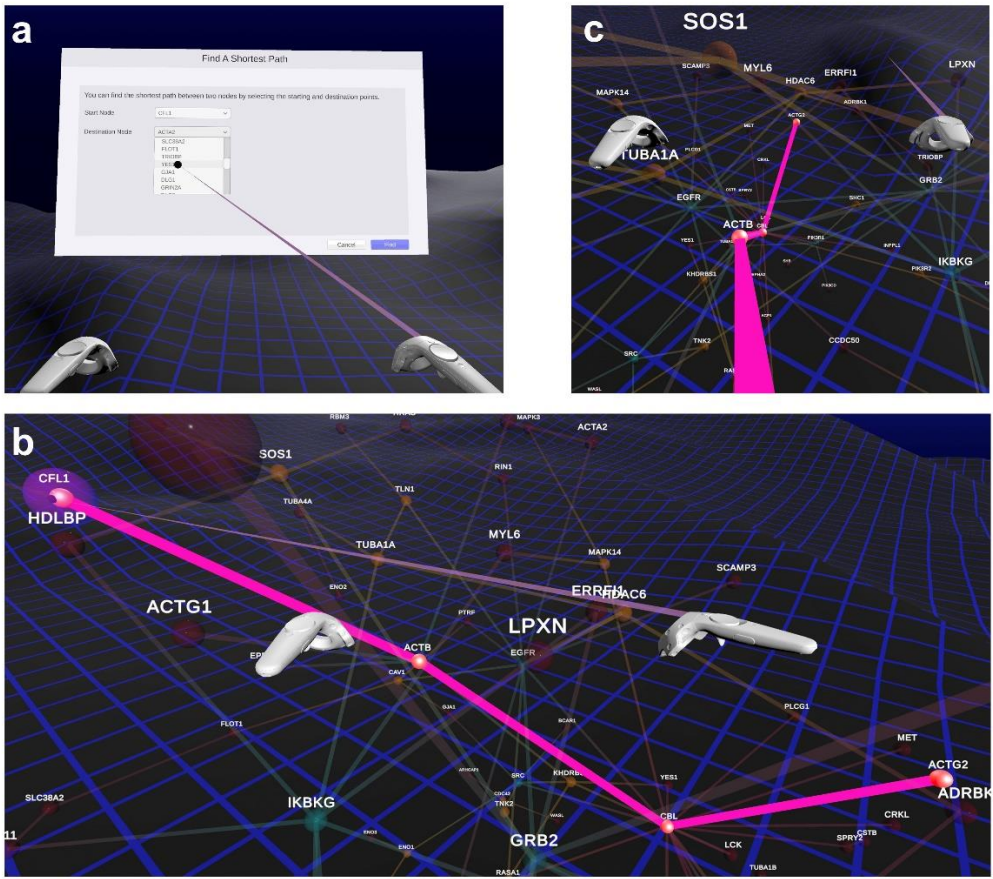

**Figure 3.** Finding the shortest path in the VR environment.

Subnetworks are smaller networks within an interactome that help identifying genes, pathways, and processes associated with a context such as diseases, drug treatments (Tuncbag, et al., 2016).

In BioNet-XR, subnetworks are used to focus on the interaction of specific nodes within the network. These nodes that the subnet is intended to contain are called seed nodes. To find a subnet, the seed nodes are determined and given as input to the algorithms. There are two ways to identify seed nodes in BioNet-XR: the user selects them from the node list or uploads a file containing the seed node names. The second method makes the selection process more straightforward for a large number of nodes. BioNet-XR includes three options for finding a subnetwork: All-Pairs Shortest Path, Steiner Tree, and Random Walk (**Figure 4**).

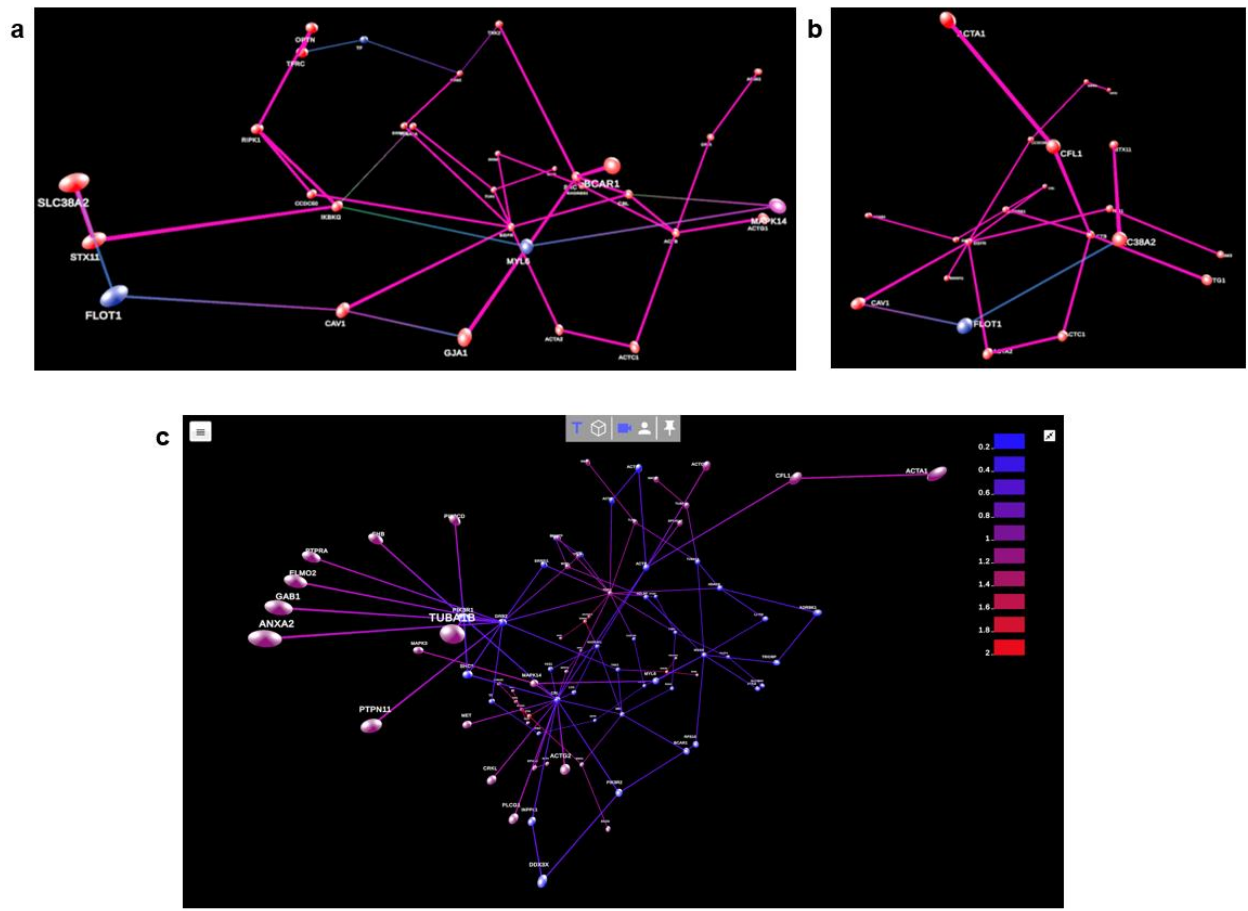

**Figure 4.** A visual demonstration of the subnetworks generated by (a) all-pairs shortest paths, (b) Steiner tree, and (c) random walk with restart algorithms.

### 3.1. MR network visualization experience with BioNet-XR

The MR version offers a unique and immersive protein network viewing experience, allowing users to interact with and explore protein networks (move, resize, and rotate) through intuitive hand gestures. While it may not include the extensive analysis functionalities found in the desktop and VR versions, it excels in providing users with the ability to examine protein networks from various angles and perspectives while maintaining a connection to the real-world environment they are observing (**Figure 5**). When users step towards the network, they will realize they are approaching it. Alternatively, when they rotate around the network, the network angle will shift in that direction. The MR version provides immersive interaction with a virtual network in the real world. When users move closer to the network, they will notice their proximity to the network. Alternatively, when they walk around the network, the network's orientation will change accordingly. The MR version offers an immersive interaction with a virtual network within the real world (see **Video S10**).

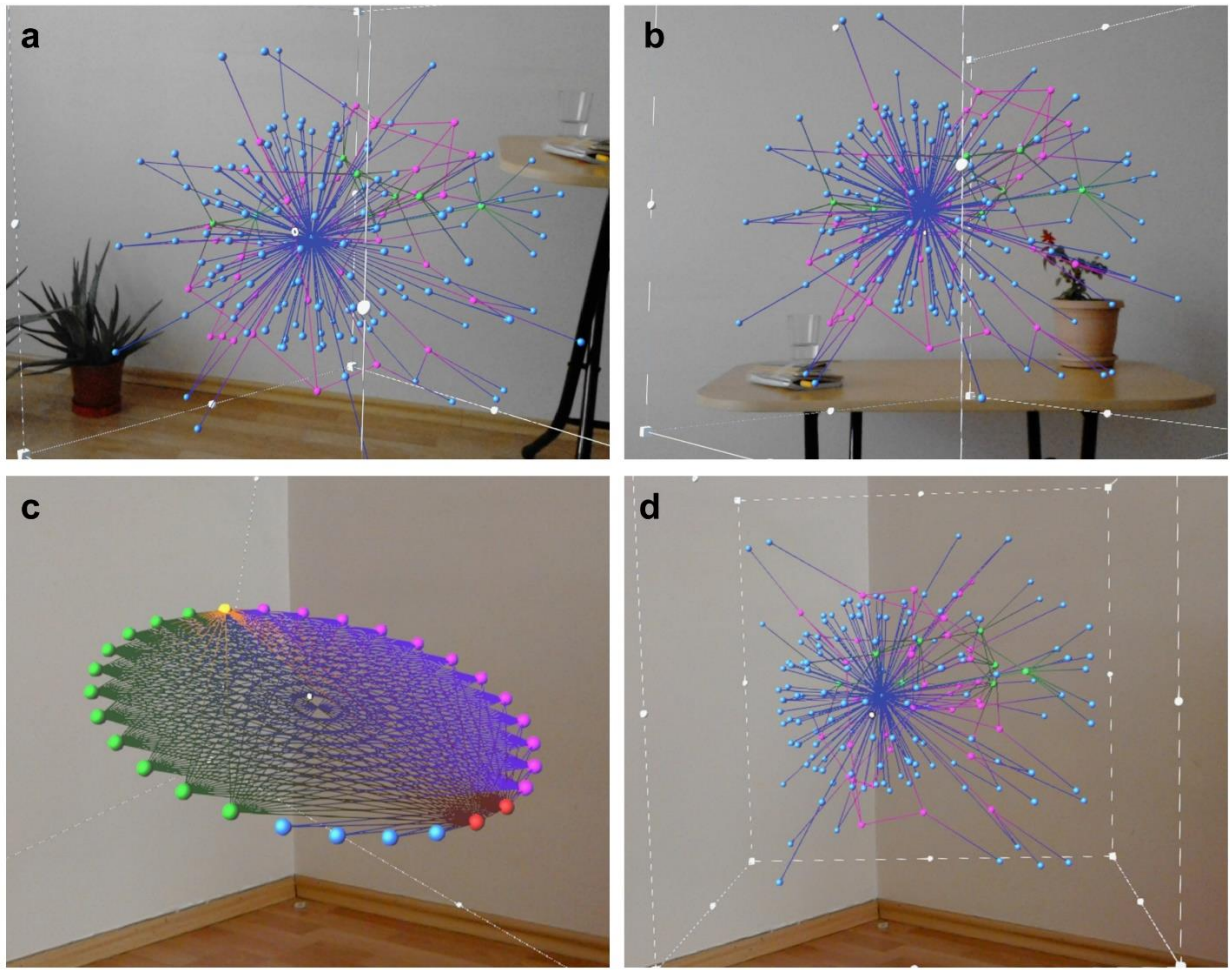

**Figure 5.** MR version of BioNet-XR, displayed on the different locations of the physical world.

### 3.2. Usability Test Results

Quantitative analysis was performed using SUS and TAM questionnaires collected from usability tests from different departments (**Figure 6a**). Responses to the open-ended questions were utilized in conducting both qualitative and quantitative analyses. Statistical analyses were performed using JASP (Goss-Sampson, 2019), an open-source statistical analysis tool.

SUS scores are calculated, resulting in an overall application usability score of 80.8. The multidisciplinary group averaged 76.3, while the Bioinformatics group achieved a higher average of 85.3, as indicated in Table 1. Usability was evaluated as an 'Excellent' rank in general and in each group.

**Table 1:** Descriptive statistics of average SUS scores of groups.

|  | Average SUS Scores | |
| --- | --- | --- |
|  | Multidisciplinary group | Bioinformatics group |
| Mean | 76.3 | 85.3 |
| Std. Deviation | 15.7 | 10.4 |
| Minimum | 45 | 62.5 |
| Maximum | 95 | 100 |

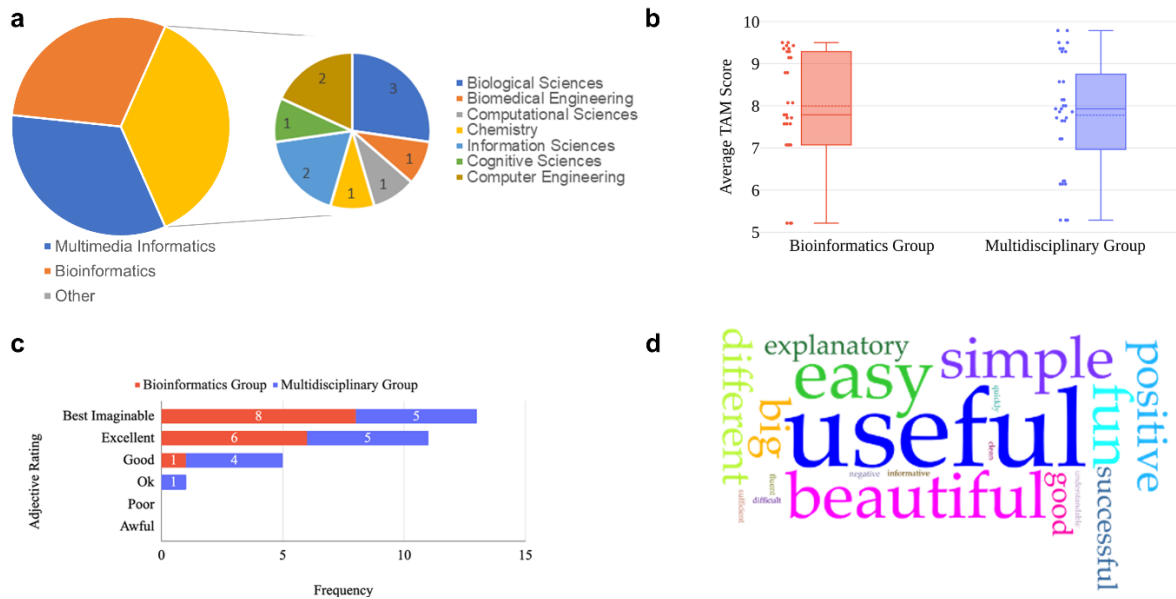

**Figure 6 (a)** Departments of the participants in usability tests **(b)** Boxplots of average TAM scores of Bioinformatics and Multidisciplinary groups **(c)** Histogram of SUS scores of groups **(d)** The word clouds are composed of the 20 most frequent adjectives in the given answers to the question about the first impression.

The TAM score was calculated by averaging the item scores (**Figure 6b**). It is important to note that the third item scores were used by subtracting the maximum scale score of 10 since it is a reversed-scaled item. Before analyzing TAM scores, reliability analysis was applied to the questionnaire to evaluate the internal consistency using Cronbach's alpha (α), with a value above 0.70 indicating acceptable reliability (Cronbach, 1951). The alpha for the total scale was 0.843, which is acceptable reliability. However, removing the third item, "During the test, I felt pain and/or discomfort," would increase reliability. Hence, it is indicated as a reversed-scaled item which means that 0 is the best score, while 10 is the worst score for the item. As a result, the absolute reliability of the scale is α = 0.886.

After validating the reliability of the questionnaire, descriptive statistics were applied. In the descriptive analysis results, the mean TAM score of the Bioinformatics group ($\mu = 8.143$) is higher than the multidisciplinary group ($\mu = 7.910$). The overall average is 8.026, which is a high score considering the maximum possible score (10).

**Figure 6c** shows the SUS score distributions by the adjective scale. Although the ratings were the same between the groups, minimal difference was observed in the score distributions. The lowest usability evaluation of the Bioinformatics group was 'Good,' while the multidisciplinary group was 'OK.' The score stack is at the highest usability rating, 'Best Imaginable,' for the Bioinformatics group. In the multidisciplinary group, scores were nearly evenly distributed among 'Good,' 'Excellent,' and 'Best Imaginable.' The result of SUS showed that BioNet-XR has good usability.

The concentration scores in the Bioinformatics group are higher than seven. The distribution is more dispersed in the multidisciplinary group, and some gave low scores. Based on a t-test, we found that the technology acceptance levels of bioinformatics and multidisciplinary groups are statistically not different from each other ($p$-value $= 0.621$).

Acceptable by the Bioinformatics group is an essential indicator for BioNet-XR to be presented to the target audience. Such software is generally challenging to appeal to users outside the domain. The absence of differences between the out-of-field and in-field groups suggests that BioNet-XR has successfully overcome this challenge.

A comparison was made between those who used a visualization tool and those who did not. Some participants from the in-field group did not use it. Data on subjects' use of the visualization tool was collected through open-ended questions. As with the previous test, there was no difference in technology acceptance levels between these groups. In other words, it is a useful and sufficient framework for those who have used similar software and those who have not.

The answers given to the question about first impressions of BioNet-XR were analyzed. Word clouds were created from the 20 most frequent adjectives, both are translated from the original Turkish answers (**Figure 6d**). The side effects of using positive adjectives in negative sentences and negative adjectives in positive sentences have been eliminated. The most frequent adjectives consist of useful, simple, easy, positive, fun, different, and simple. They are all positive as semantic. These positive words for first impressions show that BioNet-XR positively affects participants.

Table S1 has been prepared from the answers about the strengths and weaknesses of BioNet-XR. Although they have not tried it, they mention VR/MR integration as its powerful feature. Additionally, being 3D is one of the strongest features mentioned. However, it is stated as a weak feature that it does not offer as wide a range of features as Cytoscape. While true, it is not a fair comparison. Because, thanks to its well-established community, Cytoscape has gained many features over the years, and new features are still being added. Considering the SUS and TAM results, our framework's weaknesses are not crucial and do not disrupt the user experience.

Participants emphasized the availability of VR and MR versions. Generally, they indicated both versions are beneficial, but there are significant negative and abstaining opinions regarding the MR version. While receiving feedback via video offers some insight, experiencing VR and MR versions is necessary for more comprehensive feedback.

Although the functions in the VR version were not available, the development of the MR version was important. Because users cannot see their real environment in VR, users may not feel safe while moving and walking around for fear of hitting the objects in the physical environment.

However, in MR, users feel much more comfortable as the real environment is visible for navigation. Also, it is easier and more efficient to collaborate with more than one user.

## 4. Conclusion

This study presents BioNet-XR, a framework for visualizing PINs available in PC, VR, and MR versions. The PC version has three layout algorithms and visual adjustment options. The layout algorithms are force-directed single circular and single circular with Louvain community detection. The user can view the network parameters, find the shortest path between two nodes, and find subnets with three algorithms: all-pairs shortest path, Steiner tree, and random walk. The VR version has similar features to the PC version, while the MR version allows importing networks created in the PC version. In MR, users can navigate but lack analysis features. The usability evaluation was made to understand how effectively users can use the BioNet-XR. Since VR and MR versions can be used with headsets, data acquisition could not be performed due to pandemics. The desktop version of the framework was tested by 30 participants from diverse backgrounds consisting of bioinformatics-related fields and other disciplines' participants. According to the results of the SUS evaluation, the overall average and in-field average correspond to excellent in the adjective rating. It shows that BioNet-XR is easy to use for participants familiar with other tools and does not force their usage habits. Also, it has a smooth learning curve for others who do not know how to use such tools. In light of user feedback and open-ended questions, BioNet-XR has left a positive impression. Being 3D, VR/MR integration, simple interface, and clear visualization are some of its powerful features. Although the existing community may be distant from the use of VR and MR technologies, considering the technological habits and tendencies of the next generation that will be added to the community, it can be predicted that they prefer the use of such tools. Therefore, researching the potential of immersive technologies in information visualization tools makes a significant contribution.